\documentclass[10pt,twocolumn,amsmath,amssymb,aps,physrev,floatfix,longbibliography]{revtex4-2}
\usepackage{graphicx}
\usepackage{dcolumn}
\usepackage{bm}
\usepackage{siunitx}
\usepackage{svg}
\usepackage[T1]{fontenc}

\usepackage{soul} 

\begin{document}

\preprint{APS/123-QED}

\title{Polarization properties of photon Bose-Einstein condensates }

\author{Sven Enns$^{1,2,3,}$}
\altaffiliation{ORCID: 0009-0000-8092-5676}
\affiliation{$^1$Physics Department and Research Center OPTIMAS, RPTU Kaiserslautern-Landau, 67663 Kaiserslautern, Germany}
\affiliation{$^2$Leibniz University Hannover, Institute of Photonic Systems and Technologies, 30167 Hannover, Germany}
\affiliation{$^3$Leibniz University Hannover, Cluster of Excellence PhoenixD (Photonics, Optics, and Engineering Innovation AcrossDisciplines), Hannover, Germany}

\author{Julian Schulz$^{1,2,3,}$}
\altaffiliation{ORCID: 0000-0003-4630-4117}
\affiliation{$^1$Physics Department and Research Center OPTIMAS, RPTU Kaiserslautern-Landau, 67663 Kaiserslautern, Germany}
\affiliation{$^2$Institute of Photonic Systems and Technologies, Leibniz University Hannover, 30167 Hannover, Germany}
\affiliation{$^3$Cluster of Excellence PhoenixD (Photonics, Optics, and Engineering Innovation AcrossDisciplines), Hannover, Germany}

\author{Kirankumar Karkihalli Umesh$^{4,5,}$}
\altaffiliation{ORCID: 0000-0002-3644-1233}
\affiliation{$^4$Institute of Applied Physics, University of Bonn, 53115 Bonn, Germany}
\affiliation{$^5$Kirchhoff-Institute for Physics, Heidelberg University, 69120 Heidelberg, Germany}

\author{Frank Vewinger${^4,}$}
\altaffiliation{ORCID: 0000-0001-7818-2981}
\affiliation{$^4$Institute of Applied Physics, University of Bonn, 53115 Bonn, Germany}

\author{Georg von Freymann$^{1,2,3,6,}$}
\altaffiliation{ORCID: 0000-0003-2389-5532}
\affiliation{$^1$Physics Department and Research Center OPTIMAS, RPTU Kaiserslautern-Landau, 67663 Kaiserslautern, Germany}
\affiliation{$^2$Institute of Photonic Systems and Technologies, Leibniz University Hannover, 30167 Hannover, Germany}
\affiliation{$^3$Cluster of Excellence PhoenixD (Photonics, Optics, and Engineering Innovation AcrossDisciplines), Hannover, Germany}
\affiliation{$^6$Fraunhofer Institute for Industrial Mathematics ITWM, 67663 Kaiserslautern, Germany.}

\date{\today}

\begin{abstract}
The first experimental realization of a photon Bose-Einstein condensate was demonstrated more than a decade ago. However, the polarization of the condensate has not been fully understood and measured in this weakly driven-dissipative system. In this article, we experimentally investigate the polarization of thermal and condensed light depending on the power and polarization of the pump beam. With full control over the polarization of the pump light, it is possible to create arbitrary states on the surface of the Poincaré sphere. We show that, in agreement with previous theoretical work, there is a remarkable increase in the polarization strength of the condensate above the threshold for a fully linearly polarized pump light, indicating that the weak drive lifts the polarization degeneracy of the system. Above a certain threshold also the degenerate orthogonal polarized state of the cavity is occupied, limiting the degree of polarization to approximately 90\%.
\end{abstract}

\maketitle

\section{\label{sec:level1}Introduction}

Bose-Einstein condensation is a quantum phase transition of a Bose gas leading to a macroscopic occupation of the system's ground state above a critical phase-space density \cite{Einstein}. While the first realization of a Bose-Einstein condensate (BEC) was achieved in atomic systems \cite{Ketterle, Cornell}, further experimental setups have been developed to investigate condensation of magnons \cite{Nikuni,Demokritov, Koster}, exciton-polaritons \cite{Kasprzak}, plasmons \cite{Hakala}, photons in a dye-filled microcavity \cite{klaers2}, optical fibers \cite{Fischer} and recently also in vertical-cavity surface-emitting lasers (VCSELs) \cite{Pieczarka} and further semiconductor cavity systems\cite{Schofield}. These experiments advance the understanding of many fields, such as quantum thermodynamics \cite{Umesh} and quantum many-body physics. Recent works on state preparation, e.g. the condensation into a hybridized ground state \cite{Redmann}, thermalization of a two-level system \cite{Kurtscheid} and the creation of continuous condensates \cite{Chen} open the way for possible applications in quantum simulation \cite{Bloch, Huang}, quantum computing \cite{Mohseni} and sensing \cite{Bennett}. In all of these areas, the polarization of the condensate can be an important property.
Studies of polarization properties in exciton-polariton systems revealed strong polarization effects above condensation threshold, both theoretically \cite{EP_Theory_1, EP_Theory_2} and experimentally \cite{EP_1, EP_2, EP_3}, where polarization strength can reach up to about 90\% under certain conditions \cite{EP_3}. Although affected by several factors, the polarization tends to be pinned to one of the crystallographic axes of the cavity \cite{EP_2, EP_3}. This results from an anisotropy in the cavity, such as birefringence, leading to an energetic splitting of the polarization modes and hence determining the polarization of the condensate. However, the pump polarization and interaction processes such as polariton-polariton scattering can lead to rotation and an increase in its circular degree of polarization \cite{EP_Theory_1, EP_1}. This is different for photons in a dye-filled cavity, where the influence of the pump radiation can be studied in the absence of birefringence. 

\section{\label{sec:theoryReview}Theory Review}
Theoretical models for photon gases in a dye-filled microcavity also predict the build-up of strong condensate polarization. While thermalization of the photon gas can also be observed spatially, where symmetry is preserved \cite{KirtonKeeling_2}, theoretical investigations by R. Moodie et al. \cite{KirtonKeeling} show that the polarization state of the condensate is not necessarily fully thermalized. By assuming strongly anisotropic dye molecules and accounting for rotational diffusion, simulations yield a strong increase in the polarization strength of the condensate above threshold for a linearly polarized pump, while the thermal cloud remains nearly unpolarized. This arises from the interplay between the timescales of molecular excitation lifetime, spontaneous and stimulated emission, and photon lifetime. Polarized pump light preferably excites molecules with their dipole moment oriented along the polarization direction. Rotational diffusion leads to a reorientation of the dipole moment before the photon gets re-emitted. For thermal modes below and above threshold, the emission timescale and spontaneous lifetime are close and full rotational diffusion is expected, leading to an unpolarized thermal cloud. For the condensate above threshold, stimulated processes faster than the rotational diffusion become more dominant and result in an increasing polarization strength of the condensate, whereas in a perfectly closed system complete thermalization prevents the appearance of polarization effects through equal occupation of the polarization degenerate ground state of the system. This emphasizes the difference of the weakly driven-dissipative system in comparison to a closed system. 

While the condensate exhibits strong polarization aligned with the pump, a second condensation threshold is observed at higher pump powers. This can be explained by the fact that most molecules couple to both polarization modes. Together with rotational diffusion, this results in emission into both polarization states. Since both polarization components of the thermal cloud are already close to threshold, the mode polarized perpendicular to the pump can also undergo condensation at higher pump powers depending on the cavity losses. This behavior is further enhanced by an effect similar to the spatial hole-burning effect known from laser physics. As excitations of molecules oriented along the pump polarization become saturated, the gain for molecules with other orientations increases, thereby promoting condensation in the orthogonal polarization mode.

The main aspects to be confirmed experimentally are:
\begin{itemize}
    \item Strong degree of polarization of the condensate for linear pump polarization 
    \item Polarization orientation is dictated by pump polarization
    \item Build up of second condensation threshold with orthogonal polarization at high enough pump power
    \item Influence of pump polarization on critical pump power
\end{itemize}

First experimental work on the polarization of photon Bose-Einstein condensates confirmed the strong increase of the linear polarization degree of the condensate \cite{vanOosten}. However, since the dye was pumped at an angle, the influence of the pump polarization remains unknown. By aligning the pump beam to the optical axis we confirm the power-dependent behavior of the condensate polarization and investigate the influence of the pump polarization as the only source of rotational symmetry breaking. Thereby, it dictates the dominant polarization state of the condensate.

The confirmation of these results would enhance the comprehension of the driven-dissipative nature of the system. This influence has previously been observed when changing the pump beam geometry \cite{Marelic}. In this instance, however, it is evident in the non-complete thermalization of the polarization. Being able to control the polarization state of the condensate via pump or designed asymmetries, possibilities especially in lattice physics arise. As coupling strengths between neighboring sites can be modified, a new timescale is introduced that can be controlled up to a certain degree. This can be a new approach for the realization of vortices or skyrmions in photonic systems.

\section{Experimental Setup}

\begin{figure}[bt!]
\includegraphics[scale=0.5]{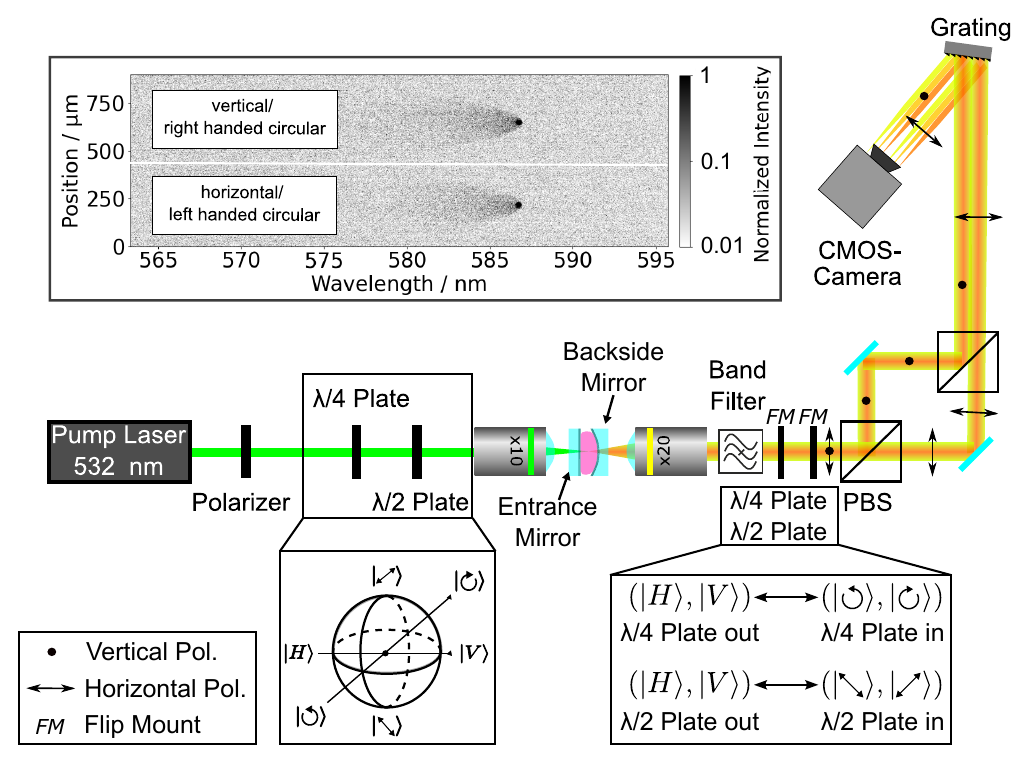}
\caption{Schematic illustration of the experimental setup. Passing a quarter-wave plate and a half-wave plate, the polarization of the pump beam can be set to an arbitrary polarization state on the surface of the Poincaré sphere. After focusing the pump beam into the dye-filled cavity the fluorescence leaving the resonator at the back-side mirror is separated into two orthogonal polarization states by a polarizing beam splitter (PBS). A quarter-wave plate or a half-wave plate is used to switch from the horizontal-vertical measurement basis to the circular or the 45°-rotated linear polarization measurement basis, respectively. The inset shows an exemplary wavelength resolved measurement of the cavity emission, where the condensate is distinguishable from the thermal cloud and the two polarization states are clearly separated.}
\label{fig:experimentalSetup}
\end{figure}

In the following, we describe the experimental setup used to analyze the polarization state. For a more detailed description of the dye-microcavity system see e.g. \cite{klaers1, klaers2}, and Appendix A gives a summary of the theoretical concepts.
The experimental setup used to analyze the polarization states of the photon gas is shown schematically in Fig. \ref{fig:experimentalSetup}. A vertically polarized quasi continuous-wave laser beam of wavelength \SI{532}{nm} is focused through the entrance mirror into a microcavity filled with a $\SI{1.5}{mmol/l}$ Rhodamine 6G solution in ethylene glycol. With the pump beam aligned along the optical axis, the setup remains rotationally symmetric. The pump polarization can be set to any point on the surface of the Poincaré sphere using a polarizer followed by half- and quarter-wave plates. Hence, the investigation can be extended to also circular pump polarization. The alignment of the pump beam along the optical axis, the rotational symmetry and the control of the pump polarization state allow the investigation of the correlation between pump and photon-gas polarization, unperturbed by other symmetry breaking effects.

The cavity consists of a plane and a concave Bragg mirror (radius of curvature $R=\SI{1}{m}$) spaced by a distance corresponding to $9\times\lambda_{\text{cut}}/2$ ($\approx$1.9\,µm) at longitudinal mode number $q=9$ with $\lambda_{\text{cut}}$ denoting the cutoff wavelength. Thereby, creating a confining parabolic potential that enables thermalization and condensation of a two-dimensional photon gas \cite{klaers1,klaers2}. Light exiting the cavity through the backside mirror passes through a band filter blocking the pump light. A polarizing beam splitter separates the horizontally and vertically polarized parts of the remaining fluorescence. Using a half-wave (quarter-wave) plate before the beam splitter, switches the horizontal-vertical measurement basis to a 45\textdegree-rotated linear polarized (circular polarized) basis. Both separated paths are parallel with a spatial offset. Imaging the light after diffraction on a grating with a CMOS camera shows the wavelength resolved intensity distribution of the photon gas, where both polarization fractions can be measured simultaneously and analyzed for each individual mode, similarly to \cite{Umesh}. The different configurations allow the measurement of the polarization state in terms of the Stokes parameters.
The inset of Fig. \ref{fig:experimentalSetup} shows an exemplary measurement of a photon gas with pump powers above condensation threshold. One observes macroscopic occupation of the ground mode (indicated by dark dots of highest intensity) at 587\,nm, followed by the thermal cloud for smaller wavelengths. Thereby, the two separated fractions of condensate and thermal cloud represent two orthogonal polarization states.


\section{Results}
First we investigate the relation between the polarization of pump and photon gas. The normalized intensities  $n_{\phi}$ for linear polarization orientation $\phi$ and $n_{\text{LH}} / n_{\text{RH}}$ for left- and right-handed circular polarization of thermal cloud and condensate are determined from the corresponding counts on the CMOS camera. The polarization state is defined by the Stokes parameters as

\begin{equation}
    S_{\text{1}} = \frac{n_{\text{0°}}-n_{\text{90°}}}
    {n_{\text{0°}}+n_{\text{90°}}},
    \label{eq:S1}
\end{equation}
\begin{equation}
    S_{\text{2}}= \frac{n_{\text{45°}}-n_{\text{135°}}}
    {n_{\text{45°}}+n_{\text{135°}}},
    \label{eq:S2}
\end{equation}
\begin{equation}
    S_{\text{3}} = \frac{n_{\text{RH}}-n_{\text{LH}}}{n_{\text{RH}}+n_{\text{LH}}},
\label{eq:S3}
\end{equation}

where the index denotes the polarization orientation. In the linearly polarized basis,  $S_{\text{1}}=-1 (+1)$ corresponds to fully horizontal (vertical) polarization, while $S_{\text{1}}=0$ is obtained for all polarizations on the equatorial plane of the Poincaré sphere. For each Stokes parameter the measurement is performed in the corresponding measurement basis by implementing a half-wave plate behind the cavity to measure $S_{\text{2}}$ and a quarter-wave plate to measure $S_{\text{3}}$.

\begin{figure*}[htb!]
\centering
\includegraphics[scale=0.52]{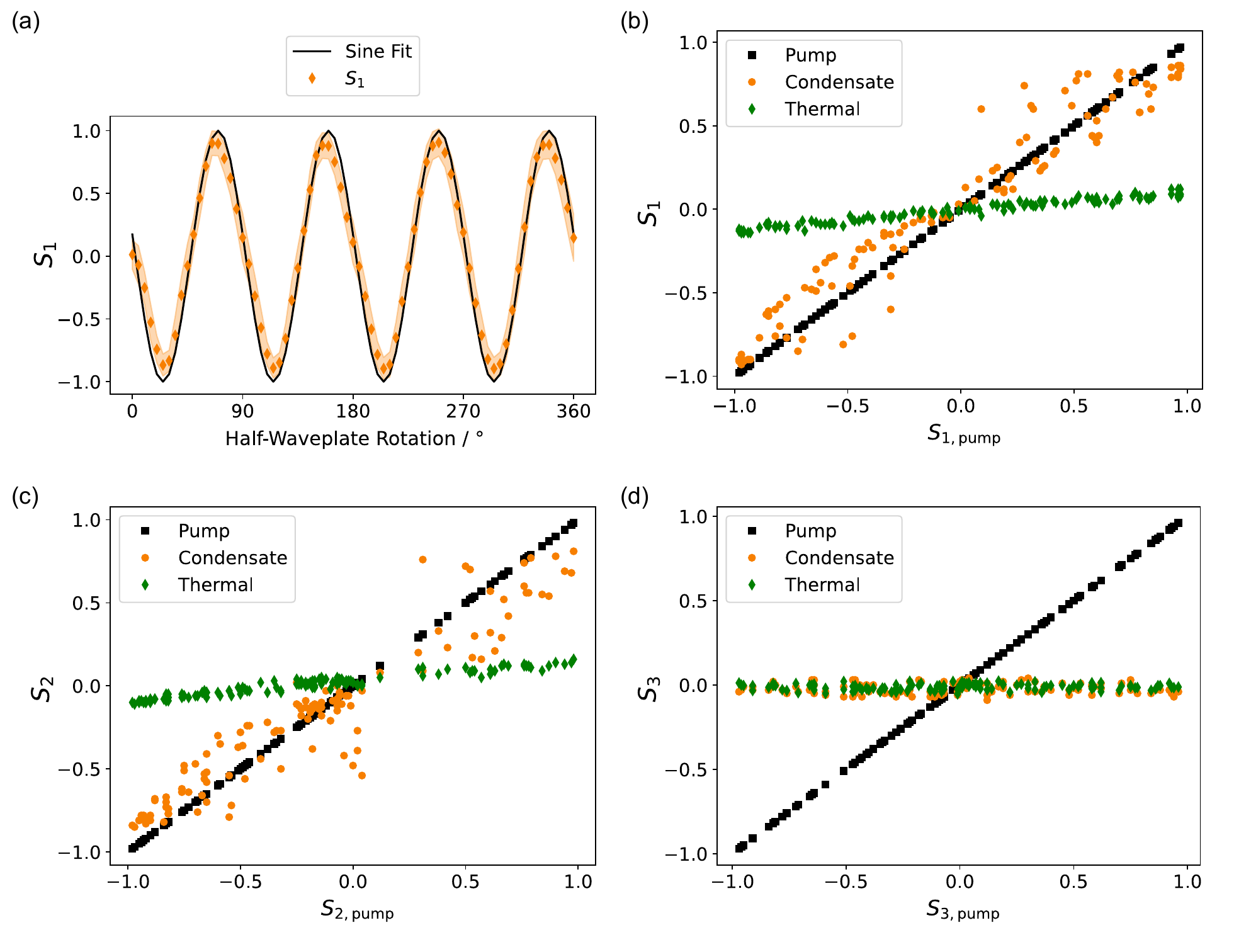}
\caption{Polarization strength of the condensate and thermal cloud in comparison to the pump polarization. (a) The rotation of the pump polarization using a half-wave plate leads to a rotation of the condensates polarization and shows the expected behavior for the polarization strength, following a sine function (solid line). Rastering the surface of the Poincaré sphere for the pump polarization leads to the resulting Stokes parameters shown in (b)-(d), each sorted from lowest to highest value of the corresponding pump Stokes parameter.}
\label{fig:PumpDependence}
\end{figure*}

To study the relation between pump and condensate polarization in the linear measurement basis of Stokes parameter $S_{\text{1}}$, a half-wave plate is used to rotate the pump polarization. The intensities of the condensate in both horizontal and vertical polarization are measured as a function of the rotation angle of the half-wave plate and show the expected behavior, so that $S_{\text{1}}$ follows a sine function, as visualized in Fig. \ref{fig:PumpDependence}(a) with an average condensate fraction of around 50\%. It is evident that the polarization of the condensate follows that of the pump beam in the linearly polarized case. The highest measured value of $S_{\text{1}}\approx 0.9$ underlines the strong polarization of the condensate.  
The measurements are extended to elliptically and circularly polarized pump light using the combination of a half-wave and a quarter-wave plate to raster the surface of the Poincaré sphere. The grid is prepared such that linear, elliptical, and circular pump polarizations with high polarization strengths are created. A reference measurement of the pump polarization is performed, to account for polarization dependent reflection and transmission effects of the cavity Bragg mirrors and the grating, imperfections of the wave plates, and other possible influences of optical components. Thereby, the measurements of each Stokes parameter for pump and photon gas are performed separately in the corresponding measurement basis in the same way as presented in Figure \ref{fig:experimentalSetup}. This allows a proper comparison between the pump and the photon gas. For all determined polarization states of the pump the polarization states of the condensate and the thermal cloud are analyzed. The comparison between the polarization states of pump (black), ground mode of the photon gas (orange) and its thermal modes (green) is shown in Figure \ref{fig:PumpDependence}(b)-(d), where for each plot the measurement datasets containing the Stokes parameters of pump, condensate and thermal cloud are sorted from lowest to highest value of the corresponding pump Stokes parameter. It is clearly evident, that for Stokes parameters $S_{\text{1}}$ and $S_{\text{2}}$ describing the linear polarized fraction a strong correlation is present. The degree of polarization of the condensate is limited to up to $P \approx 0.9$, as the thermal part in the ground mode cannot be separated from the condensate. Nevertheless, the data confirm the theoretical results, where the polarization state of the condensate is dictated by the pump in the linear polarized case. The thermal cloud shows only weak polarization effects. A continuous increase of polarization strength indicates a correlation with the pump polarization. However, for strongly linear polarized pump $S_{\text{1,pump}}\approx S_{\text{2,pump}}\approx \pm 1$ the thermal cloud shows maximum values of $S_{\text{1,thermal}}\approx S_{\text{2,thermal}}\approx \pm0.15$. Here, predominantly low-energy modes contribute, as they are strongly populated and stimulated emission can not be neglected. In contrast to the linearly polarized fraction, no polarization effects are visible for the circularly polarized fraction of condensate and thermal cloud, as the measurements of $S_{\text{3}}$ illustrate.

\begin{figure}[tb!]
\centering
\includegraphics[scale=0.52]{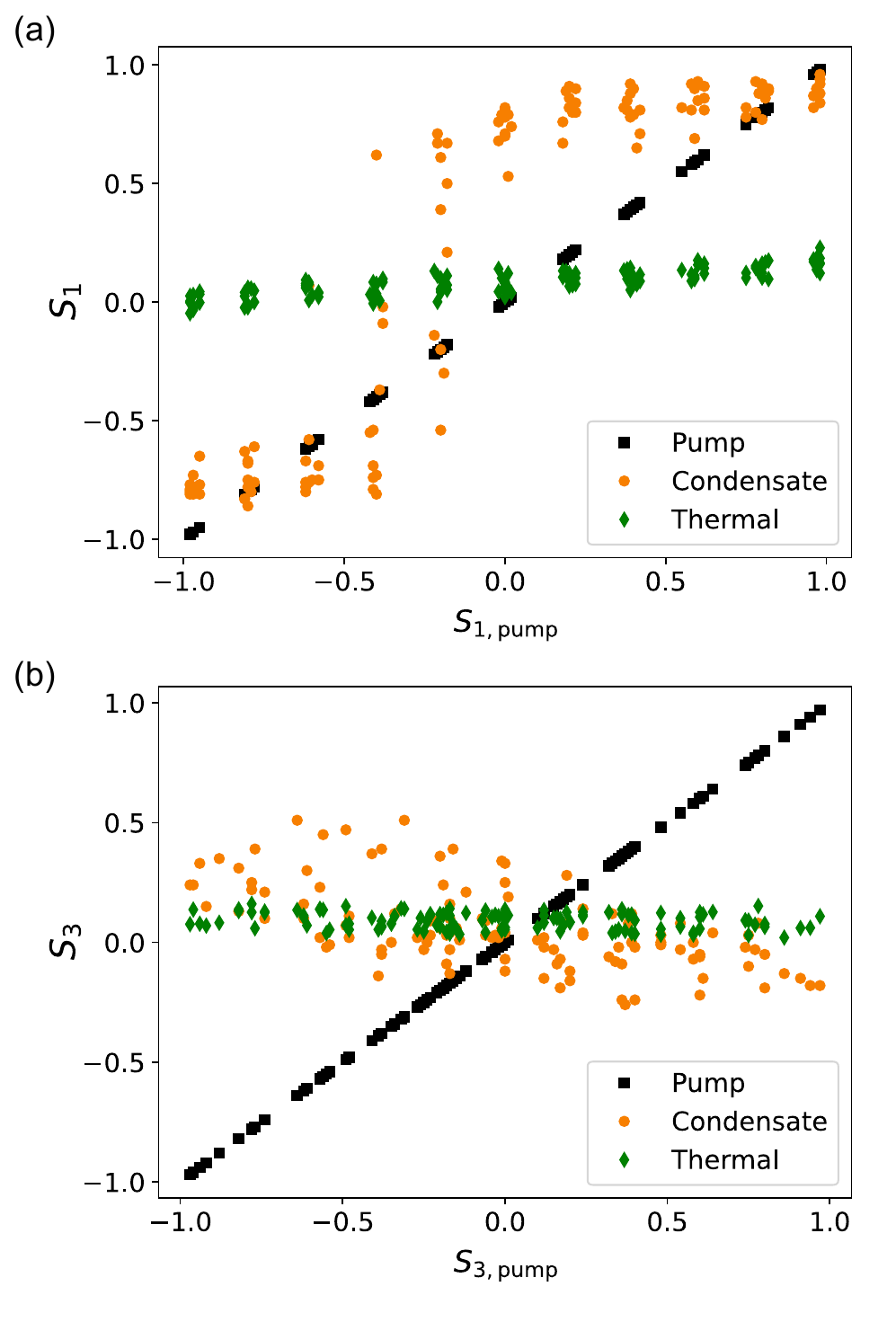}
\caption{Polarization strength for $S_{\text{1}}$ in (a) and $S_{\text{3}}$ in (b) of condensate and thermal cloud compared to the pump polarization. Demonstration of polarization pinning effect for the condensate introduced by defects in the Bragg layers of the mirror, creating an asymmetry in the setup. Thermal cloud still shows only small polarization effects, not influenced by the asymmetry.}
\label{fig:Pinning}
\end{figure}

To interpret the results, emission properties of the dye molecules have to be taken into account. Rhodamine 6G is well suited to investigate polarization in this setup, since dipole moments of absorption and emission stay nearly parallel \cite{Zhou, Lopez, Penzkofer}, leading to an unperturbed insight to the thermalization of polarization properties, determined by the timescales of rotational diffusion of the molecules and excitation lifetime. As explained, stimulated emission above threshold is faster than rotational diffusion, thus the good agreement between pump and condensate polarization can be explained by the nearly parallel dipole moments. Furthermore, as Rhodamine 6G is a strongly anisotropic molecule with a dominant and non-degenerate dipole moment, circular polarized emission is not expected, since two dipole transitions with fixed phase relation would be needed. Additionally, the multiple absorption and emission cycles of photons inside of the cavity can destroy possible coherence of the orthogonal polarized condensates, preventing circular polarization effects to occur. Although there is ongoing research on how to create circularly polarized laser light from achiral dyes, the effects on the total degree of circular polarization are rather small \cite{anisotropy_1} and usually require modeling of the environment or molecular structure \cite{anisotropy_1, Nanasi}. Considering the timescales for the much faster rotation of the electrical field vector $<\SI{2}{fs}$ in comparison to one experimental realization $\approx\SI{500}{ns}$, circular polarized pumping can equally excite molecules of all possible orientations. It can therefore also be seen as unpolarized pumping, explaining the measured results of $S_{\text{3}} \approx 0$ for thermal cloud and condensate. In terms of polarization purity $\Pi = \sqrt{S_{\text{1}}^2+S_{\text{2}}^2+S_{\text{3}}^2}$, highest purity is achieved for linearly polarized pump and drops to $\Pi\approx0$ for circular polarization (see Appendix~\ref{App.Purity}). In between, elliptical polarization of the pump introduces a preferred direction and increased purity. 

While in this symmetrical case the pump dictates the preferred direction, this does not have to be the case for symmetry broken setups. 
Introducing asymmetries, e.g. due to defects in the first layers of a cavity Bragg mirror or mechanical stress, can affect the occupation of polarization states, exemplary shown for a mirror with defects in the Bragg layers in Fig. \ref{fig:Pinning}. The polarization states of the condensate clearly shift toward Stokes parameters of $S_{\text{1}}=\pm1$, aligning along or perpendicular to the defect. Depending on the asymmetry, it affects the polarization dependent transmission and reflection properties of the cavity and can lead to stronger spreading of the condensate polarization states along the circular polarization axis. Nevertheless, the polarization of the thermal cloud is not affected.
  
\begin{figure}[tb!]
\centering
\includegraphics[scale=0.335]{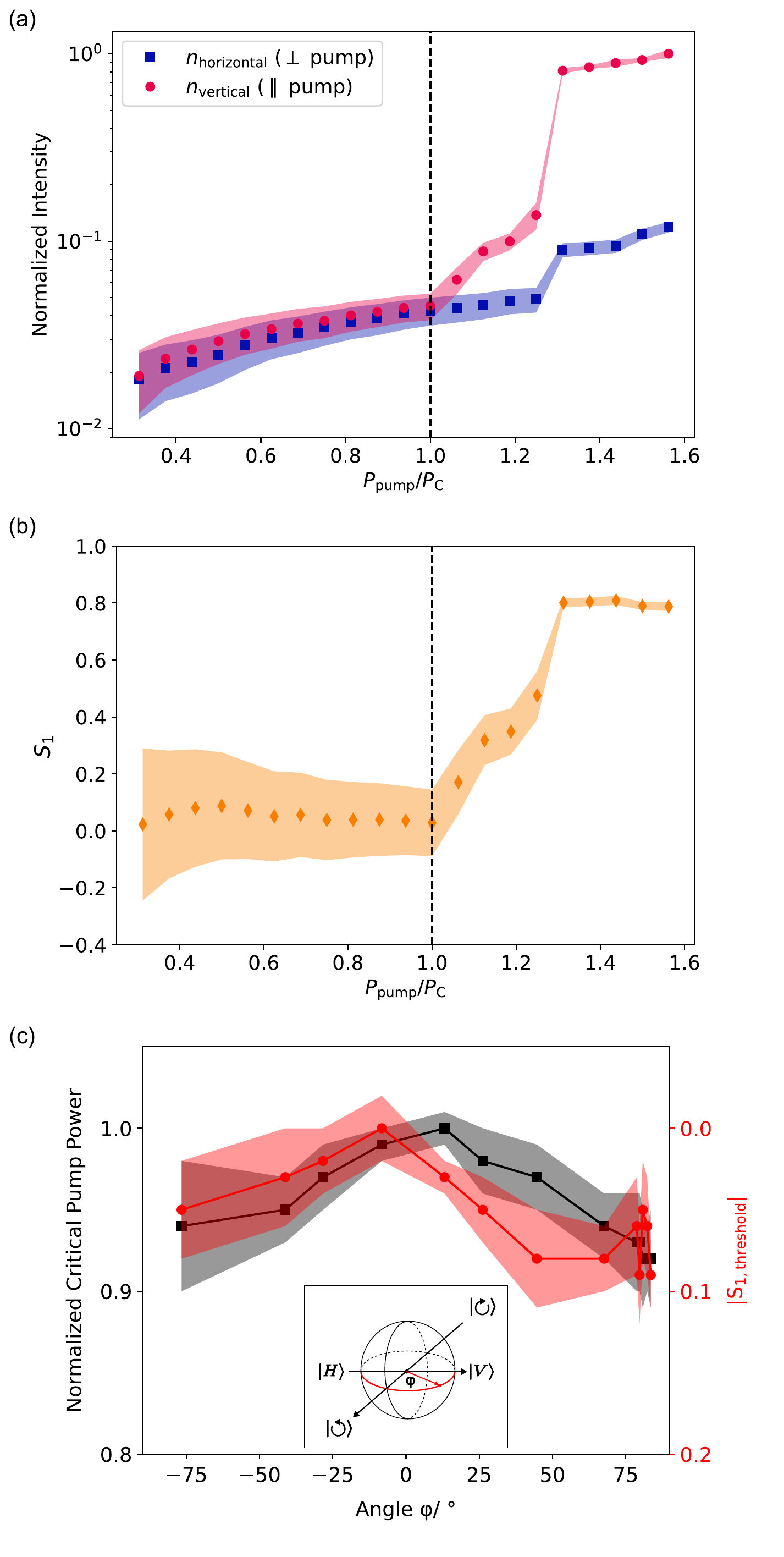}
\caption{Polarization dependence on pump power. Intensity distribution and polarization strength of the photon gas for increasing pump power with vertical pump polarization ($S_{\text{1}}\approx1$). The pump power $P_{\text{pump}}$ is shown in relation to the critical power $P_{\textbf{C}}$, where the condensation threshold is visualized by the dashed line.(a) The normalized intensities for horizontally and vertically polarized fractions of condensate and thermal cloud are shown. Each polarization is measured separately with an EMCCD camera to increase the signal-to-noise ratio and visibility of the second threshold. (b) The corresponding values of $S_{\text{1}}$ given the intensities at (a) are plotted. (c) Analysis of the critical pump power for varied pump polarization is shown in terms of angle $\varphi$. Highest critical pump power is needed for circular pump polarization, whereas a reduced critical pump power is seen for stronger linear degree of polarization. The absolute value of $S_{\text{1}}$ of the thermal cloud below threshold shows similar behavior and leads to the varied critical pump power. The markers showing the average value and the shaded areas represent the standard deviation.}
\label{fig:PowerDependence}
\end{figure}

In addition to the pump polarization also the pump power is expected to play an important role regarding the polarization of the photon gas as this influences the time scale for stimulated emission. Measuring in the linear basis of $S_{\text{1}}$ for increasing pump power with polarization of $S_{\text{1}}\approx1$ yields the results in Fig.~\ref{fig:PowerDependence}. The upper panel shows the normalized intensities of the photon gas with vertical and horizontal polarization. At the condensation threshold, vertically polarized emission increases strongly. However, further increase of the pump power also leads to the appearance of a second condensation threshold for the degenerate horizontal polarization state. This behavior has also been theoretically described in reference \cite{KirtonKeeling}, where reducing cavity loss rates can lead to condensation into the orthogonal polarization orientation. The different power thresholds are a consequence of unequally pumping both polarization orientations. Although small polarization effects can be seen in the thermal cloud for linear polarized pump as demonstrated in Fig. \ref{fig:PumpDependence} also the orthogonal component is close to condensation threshold.
Above threshold, the pump-favored orientation dominates. Because not only molecules with parallel oriented dipole moment can be excited and reorientation processes still take place, emission in the orthogonal polarization is giving rise to a second threshold at higher pump powers. Based on these measurements, Fig. \ref{fig:PowerDependence}(b) shows the calculation of the polarization strength in terms of $S_{\text{1}}$ according to Eq.~\ref{eq:S1}. The polarization strength of the photon gas below threshold is close to 0, confirming the small polarization effects, as visible in Figure \ref{fig:PumpDependence}. Above threshold, the expected strong increase of the polarization strength can be observed. The experimental data show a maximum polarization strength of $S_{\text{1}} \approx 0.8$ and do not reach $S_{\text{1}} = 1$ due to condensation into both polarization states and the contribution of the unpolarized thermal cloud.\\
Although the condensate only shows linear polarization effects, circularly polarized pumping is expected to result in higher critical pump power as both degenerate polarization states of the ground mode are occupied equally. As already indicated by the small polarization effects of the thermal cloud, linearly polarized excitation shows a favored orientation that can reach threshold at lower pump powers. To investigate this behavior, the pump polarization state is changed along the equator of the Poincaré sphere, where the current polarization state vector is defined by its angle $\varphi$ to the circular polarization axis, such that $\varphi=0$\textdegree~describes circular and $\varphi=\pm 90$\textdegree~vertical and horizontal polarization.
As shown in Fig. \ref{fig:PowerDependence}(c), the expected increase from the lowest critical power for linearly to the highest critical pump power for circularly polarized pump is visible, showing a maximum difference of around 10\%. 
Thereby, the critical pump power drops continuously for increasing degree of linear polarization.
Furthermore, the absolute value of polarization strength of the thermal cloud right before reaching the condensation threshold is plotted in red. Like the critical pump power, the polarization strength of the thermal cloud is close to 0 for circular polarized pump and is increased with higher degree of linear polarization. The polarization strength quantifies the relative difference in population of one state over the other. The normalized critical pump power and the polarization effects of the thermal cloud show the same scaling with changed pump polarization. As one of the polarization states is already slightly more populated in the thermal cloud below threshold, condensation can set in at a lower critical pump power for this polarization.\\

\section{Conclusion}
\label{sec:conclusion}
We investigated the polarization properties of a photon gas in a dye-filled microcavity below and above the condensation threshold by separating two orthogonally polarized fractions of the photon gas in linearly and circularly polarized measurement bases. Thereby, theoretical investigations could be confirmed qualitatively. Our measurements reveal the appearance of novel polarization effects due to the interplay of timescales of rotational diffusion, thermalization time and photon lifetime in the cavity. Below threshold, rotational diffusion is faster than emission processes, resulting in only small polarization effects of the photon gas. Above the threshold, when stimulated emission becomes the leading process, the condensate can show a preferred polarization orientation among the two degenerate states, governed by the polarization of the pump. Since in a perfectly closed system complete thermalization would lead to an equal occupation of the degenerate states, the appearance of strong polarization strengths emphasizes the importance of open system dynamics in the investigated dye-microcavity based setup. The polarization properties can be used to engineer coupling strength between separated condensates. Therefore, a new timescale is introduced to facilitate further study of thermalization and open dissipative systems.

\begin{acknowledgments}
We acknowledge fruitful discussions with A. Pelster, J. Krauß and M. Weitz.
We also acknowledge support by the Deutsche Forschungsgemeinschaft (DFG, German Research Foundation) under Germany’s Excellence Strategy within the Cluster of Excellence PhoenixD (EXC 2122, Project ID 390833453), through CRC/Transregio 185 OSCAR (project No. 277625399) and by the Max Planck Graduate Center with the Johannes Gutenberg University of Mainz (MPGC). 
\end{acknowledgments}

\bibliography{apssamp.bib}

@PREAMBLE{
 "\providecommand{\noopsort}[1]{}" 
 # "\providecommand{\singleletter}[1]{#1}%" 
}

@ARTICLE{Einstein,
   author       = "A. Einstein", 
   journal      = "Sitzungsber. Preuss. Akad. Wissensch.", 
   title        = "Quantentheorie des einatomigen idealen Gases. Zweite Abhandlung.", 
   year         = "1925", 
   volume       = "1", 
   pages        = "3–14",
}

@ARTICLE{Ketterle,
    author       = "K. B. Davis and M. O. Mewes and M. R. Andrews and N. J. van Druten and D. S. Durfee and D. M. Kurn and W. Ketterle", 
   journal      = "Phys. Rev. Lett.", 
   title        = "Bose-Einstein Condensation in a Gas of Sodium Atoms", 
   year         = "1995", 
   volume       = "75", 
   pages        = "3969",
}

@ARTICLE{Cornell,
   author       = "M. Anderson and J. Ensher and M.Matthews and C. Wieman and E. Cornell", 
   journal      = "Science", 
   title        = "Observation of Bose–Einstein condensation in a dilute atomic vapor", 
   year         = "1995", 
   volume       = "269", 
   pages        = "198-201",
}

@ARTICLE{Nikuni,
   author       = "T. Nikuni and M. Oshikawa and A. Oosawa and H. Tanaka", 
   journal      = "Phys. Rev. Lett.", 
   title        = "Bose-Einstein Condensation of Dilute Magnons in TlCuCl3", 
   year         = "2000", 
   volume       = "84", 
   pages        = "5868",
}

@ARTICLE{Demokritov,
   author       = "S. O. Demokritov and V. E. Demidov and O. Dzyapko and G. A. Melkov and A. A. Serga and B. Hillebrands and A. N. Slavin ", 
   journal      = "Nature", 
   title        = "Bose–Einstein condensation of quasi-equilibrium magnons at room temperature under pumping.", 
   year         = "2006", 

   volume       = "443", 
   pages        = "430-433",
}

@article{Koster,
author = {Koster, M. and Schweizer, M. R. and Noack, T. and Vasyuchka, V. I. and Bozhko, D. A. and Hillebrands, B. and Weiler, M. and Serga, A. A. and von Freymann, G.},
title = {Emergence of phase coherence in a magnon Bose–Einstein condensate},
journal = {Nat. Phys.},
volume = {},
number = {},
pages = {},
year = {2026},
doi = {https://doi.org/10.1038/s41567-026-03373-6},
}

@ARTICLE{Kasprzak,
   author       = "J. Kasprzak and M. Richard and S. Kundermann and A. Baas and P. Jeambrun and J. M. J. Keeling and F. M. Marchetti and M. H. Szymańska and R. André and J. L. Staehli and V. Savona and P. B. Littlewood and B. Deveaud and Le Si Dang ", 
   journal      = "Nature", 
   title        = "Bose–Einstein condensation of exciton polaritons.", 
   year         = "2006", 

   volume       = "443", 
   pages        = "409-414",
}

@ARTICLE{Hakala,
   author       = "T. K. Hakala and A. J. Moilanen and A. I. Väkeväinen and R. Guo and J. Martikainen and K. S. Daskalakis and H. T. Rekola and A. Julku and P. Törmä",
   journal      = "Nat. Phys.", 
   title        = "Bose–Einstein condensation in a plasmonic lattice.", 
   year         = "2018", 
   volume       = "14", 
   pages        = "739–744",
}

@ARTICLE{klaers2,
   author       = "J. Klaers and J. Schmitt and F. Vewinger and M. Weitz",
   year         = "2010",
   title        = "\text{Bose-Einstein} condensation of photons in an optical microcavity",
   journal      = "Nature",
   volume       = "468",
   pages        = "545-548",
}

@ARTICLE{Fischer,
   author       = "R. Weill and A. Bekker and B. Levit and B. Fischer", 
   journal      = "Nature Communications", 
   title        = "Bose–Einstein condensation of photons in an erbium–ytterbium co-doped fiber cavity", 
   year         = "2019", 
   volume       = "10", 
   pages        = "747",
}

@ARTICLE{Pieczarka,
   author       = "M. Pieczarka and M. Gębski and A. N. Piasecka and J. A. Lott and A. Pelster and M. Wasiak and T. Czyszanowski", 
   journal      = "Nature Photonics", 
   title        = "Bose–Einstein condensation of photons in a vertical-cavity surface-emitting laser", 
   year         = "2024", 
   volume       = "18", 
   pages        = "1090-1096",
}

@article{Schofield,
  title = {Bose–Einstein condensation of light in a semiconductor quantum well microcavity},
  author = {Schofield, Ross C. and Fu, Ming and Clarke, Edmund and Farrer, Ian and Trapalis, Aristotelis and Dhar, Himadri S. and Mukherjee, Rick and Severs Millard, Toby and Heffernan, Jon and Mintert, Florian and Nyman, Robert and A. Oulton, Rupert F.},
  journal = {Nature Photonics},
  volume = {18},
  issue = {10},
  pages = {1083-1089},
  year = {2024},
}

@ARTICLE{Umesh,
   author       = "K. \text{Karkihalli Umesh} and J. Schulz and J. Schmitt and M. Weitz and G. von Freymann and F. Vewinger", 
   journal      = "Nat. Phys.", 
   title        = "Dimensional crossover in a quantum gas of light", 
   year         = "2024", 
   volume       = "20", 
   pages        = "1810-1815",
}

@ARTICLE{Redmann,
   author       = "A. Redmann and C. Kurtscheid and N. Wolf and F. Vewinger and J. Schmitt and M. Weitz",
   journal      = "Phys. Rev. Lett.", 
   title        = "Bose-Einstein Condensation of Photons in a Four-Site Quantum Ring", 
   year         = "2024", 
   volume       = "133", 
   pages        = "093602",
}

@ARTICLE{Kurtscheid,
   author       = "C. Kurtscheid and A. Redmann and F. Vewinger and J. Schmitt and M. Weitz",
   journal      = "arXiv:2411.14838v1", 
   title        = "Thermodynamics and State Preparation in a Two-State System of Light", 
   year         = "2024", 
}

@ARTICLE{Chen,
   author = "C.-C. Chen and R. González Escudero and J. Minář and B. Pasquiou and S. Bennetts and F. Schreck",
   journal      = "Nature", 
   title        = "Continuous Bose–Einstein condensation", 
   year         = "2022", 
   volume       = "606", 
   pages        = "683-687",
}

@ARTICLE{Bloch,
   author = "I. Bloch and J. Dalibard and S. Nascimbène",
   journal      = "Nature Phys.", 
   title        = "Quantum simulations with ultracold quantum gases", 
   year         = "2012", 
   volume       = "8", 
   pages        = "267–276",
}

@ARTICLE{Huang,
   author = "Y. Huang and S. Nagata and J. Jachinowski and J. Hu and C. Chin",
   journal      = "Phys. Rev. Research", 
   title        = "Two-dimensional arrays of Bose-Einstein condensates: Interference and stochastic collapse dynamics", 
   year         = "2024", 
   volume       = "6", 
   pages        = "043272",
}

@ARTICLE{Mohseni,
   author = "M. Mohseni and V. I. Vasyuchka and V. S. L’vov and A. A. Serga and B. Hillebrands",
   journal      = "Commun. Phys.", 
   title        = "Classical analog of qubit logic based on a magnon Bose–Einstein condensate", 
   year         = "2022", 
   volume       = "5", 
   pages        = "196",
}

@ARTICLE{Bennett,
   author = "R. Bennett and D. Steinbrecht and Y. Gorbachev and S. Y. Buhmann ",
   journal      = "Phys. Rev. Applied", 
   title        = "Symmetry Breaking in a Condensate of Light and its Use as a Quantum Sensor", 
   year         = "2020", 
   volume       = "13", 
   pages        = "044031",
}

@ARTICLE{EP_Theory_1,
   author       = "D. Read and T.C.H. Liew and Y.G. Rubo and A. V. Kavokin", 
   title        = "Stochastic polarization formation in exciton-polariton Bose-Einstein condensates", 
   journal      = "PHYSICAL REVIEW B", 
   volume       = "80",
   pages        = "195309",
   year         = "2009", 
}

@ARTICLE{EP_Theory_2,
   author       = "F. P. Laussy and I. A. Shelykh and G. Malpuech and A. Kavokin", 
   title        = "Effects of \text{Bose-Einstein} condensation of exciton polaritons in microcavities on the polarization of emitted light", 
   journal      = "PHYSICAL REVIEW B", 
   volume       = "73",
   pages       = "035315",
   year         = "2006", 
}

@ARTICLE{EP_1,
   author       = "D. N. Krizhanovskii and D. Sanvitto and I. A. Shelykh and M. M. Glazov and G. Malpuech and D. D. Solnyshkov and A. Kavokin and S. Ceccarelli and M. S. Skolnick and J. S. Roberts", 
   title        = "Rotation of the plane of polarization of light in a semiconductor microcavity", 
   journal      = "PHYSICAL REVIEW B 73", 
   volume       = "73",
   pages       = "073303", 
   year         = "2006", 
}

@ARTICLE{EP_2,
   author       = "J. Kasprzak and R. André and Le Si Dang and I. A. Shelykh and A. V. Kavokin and Y. G. Rubo and K. V. Kavokin and G. Malpuech", 
   title        = "Build up and pinning of linear polarization in the Bose condensates of exciton polaritons", 
   journal      = "PHYSICAL REVIEW B",
   volume       = "75",
   pages       = "045326",
   year         = "2007", 
}

@ARTICLE{EP_3,
   author       = "\text{Ł}. Kłopotowski and M.D. Martín and A. Amo and L. Vi$\tilde{\text{n}}$a and I.A. Shelykh and M.M. Glazov and G. Malpuech and A.V. Kavokin and R. André", 
   title        = "Optical anisotropy and pinning of the linear polarization of light in semiconductor microcavities", 
   journal      = "Solid State Communications",
   volume       = "139",
   pages        = "511-515",
   year         = "2006", 
}

@ARTICLE{KirtonKeeling_2,
   author = "J. Keeling and P. Kirton",
   journal      = "Phys. Rev. A", 
   title        = "Spatial dynamics, thermalization, and gain clamping in a photon condensate", 
   year         = "2016", 
   volume       = "93", 
   pages        = "013829",
}

@ARTICLE{KirtonKeeling,
   author       = "Ryan I. Moodie and Peter Kirton and Jonathan Keeling", 
   title        = "Polarization dynamics in a photon Bose-Einstein condensate", 
   journal      = "Phys. Rev. A", 
   volume       = "96", 
   pages       = "043844",
   year         = "2017", 
}

@ARTICLE{vanOosten,
   author       = "S. Greveling and F. van der Laan and H. C. Jagers and D. van Oosten", 
   title        = "Polarization of a Bose-Einstein Condensate of Photons in a Dye-Filled Microcavity", 
   journal      = "arXiv:1712.08426v1" , 
   year         = "2017", 
}

@article{Marelic,
  title = {Experimental evidence for inhomogeneous pumping and energy-dependent effects in photon Bose-Einstein condensation},
  author = {Marelic, J. and Nyman, R. A.},
  journal = {Phys. Rev. A},
  volume = {91},
  issue = {3},
  pages = {033813},
  numpages = {5},
  year = {2015},
  month = {Mar},
  publisher = {American Physical Society},
}

@ARTICLE{klaers1,
   author       = "J. Klaers and F. Vewinger and M. Weitz",
   year         = "2010",
   title        = "Thermalization of a two-dimensional photonic gas in a 'white wall' photon box",
   journal      = "Nature Physics",
   volume       = "6",
   pages        = "512-515",
}

@article{Zhou,
author = {Zhou, Zhenpeng and Yan, Xin and Lai, Yin-Hung and Zare, Richard N.},
title = {Fluorescence Polarization Anisotropy in Microdroplets},
journal = {The Journal of Physical Chemistry Letters},
volume = {9},
number = {11},
pages = {2928-2932},
year = {2018},
}

@article{Lopez,
author = {López Arbeloa, F. and Martínez Martínez, V.},
title = {Orientation of Adsorbed Dyes in the Interlayer Space of Clays. 2 Fluorescence Polarization of Rhodamine 6G in Laponite Films},
journal = {Chemistry of Materials},
volume = {18},
number = {6},
pages = {1407-1416},
year = {2006},
doi = {10.1021/cm051518a},
}

@article{Penzkofer,
title = {Orientation of transition dipole moments of rhodamine 6G determined by excited state absorption},
journal = {Optics Communications},
volume = {35},
number = {1},
pages = {81-86},
year = {1980},
author = {A. Penzkofer and J. Wiedmann},
}

@ARTICLE{anisotropy_1,
   author       = "L. Cerdán and S. García-Moreno and A. Costela and I. García-Moreno and S. de la Moya", 
   title        = "Circularly polarized laser emission induced in isotropic and achiral dye systems", 
   journal      = "Scientific Reports", 
   volume       = "6",
   pages       = "28740",
   year         = "2016", 
}

@article{Nanasi,
author = {Nánási, Dalma E. and Johansen, Magnus B. and Mikkelsen, Kurt V. and Laursen, Bo W.},
title = {Twisting Rhodamine--Design of Bright Dyes for Circularly Polarized Fluorescence},
journal = {Journal of the American Chemical Society},
volume = {148},
number = {15},
pages = {16290-16298},
year = {2026},
}

@ARTICLE{KenStep_1,
   author       = "E. H. Kennard",
   title        = "On The Thermodynamics of Fluorescence",
   journal      = "Phys. Rev.",
   volume       = "11", 
   pages        = "29",
   year         = "1918",
}

@ARTICLE{KenStep_2,
   author       = "E. H. Kennard",
   title        = "On the Interaction of Radiation with Matter and on Fluorescent Exciting Power",
   journal      = "Phys. Rev.",
   volume       = "28", 
   pages        = "672",
   year         = "1926",
}

@ARTICLE{KenStep_3,
   author       = "B. I. Stepanov",
   title        = "Universal relation between the absorption spectra and luminescence spectra of complex molecules",
   journal      = "Dokl. Akad. Nauk SSSR",
   volume       = "112", 
   pages        = "839",
   year         = "1957",
}
\clearpage

\appendix
\section{Theoretical Introduction}
\label{App.Theory}
One of the most crucial parts for the realization of a photon Bose-Einstein condensate is the thermalization process. While thermalized photon gases naturally exist, for instance in the form of black-body radiation, the formation of a Bose-Einstein condensate (BEC) requires a conservation of the (average) photon-number in order to reach the critical photon number and to observe the macroscopic occupation of the energetic ground state. This can be achieved inside of a dye-filled microcavity consisting of two Bragg mirrors with a reflectivity around 99.998\% \cite{klaers1}, where the dye-solution is held inside of the cavity due to capillary forces. The absorption $\alpha(\omega)$ and emission $f(\omega)$ spectra of this dye solution show temperature dependent behavior and are connected by the Boltzmann constant $k_{\text{B}}$ according to the Kennard-Stepanov relation $f(\omega)/\alpha(\omega)\propto \omega^3 \text{exp}(-\hbar\omega/k_{\text{B}}T)$ \cite{KenStep_1, KenStep_2, KenStep_3}. By pumping the dye with a laser far blue detuned to the energetic ground state of the cavity, the internal energetic levels are excited and photons are emitted into the cavity. As the rovibrational states of the dye molecules equilibrate on very small time scales due to collisions with the solvent, the dye solution remains in thermal equilibrium at room temperature. Thereby the dye molecules cannot show Bose-Einstein condensation at room temperature and serve only as heat and particle bath for the photon gas. Trapping the photon gas in the resonator gives the photons a lifetime long enough to undergo several absorption and emission cycles. Since the cavity is not tuned to resonance with a specific molecular transition, an overlap between transversal cavity modes and emission spectrum of the dye is created. Additionally, this emission is incoherent since the collision processes in the dye solution also result in a randomization of the molecule orientation while the concentration ($\SI{1.5}{mmol/l}$) is too low to show interactions between the rhodamine 6G molecules. The coupling to the equilibrated heat and particle reservoir allows the thermalization of the photon gas. By choosing the mirror distance of the cavity such that photons can only be emitted into a single longitudinal mode, the photon gas becomes effectively two dimensional with the $\text{TEM}_{00}$ mode as energetic ground state of the system and higher transversal modes are occupied according to the Bose-Einstein statistics. In addition, the cut-off energy of the cavity resulting from the introduced ground mode is several orders of magnitude higher than the thermal energy, leading to vanishing thermal excitations of the dye molecules. The formation of the condensate can then be achieved above a critical phase-space density by increasing the pump power and hence the photon density \cite{klaers2}.

\section{Polarization Purity}
\label{App.Purity}
\begin{figure}[b!]
\includegraphics[scale=0.37]{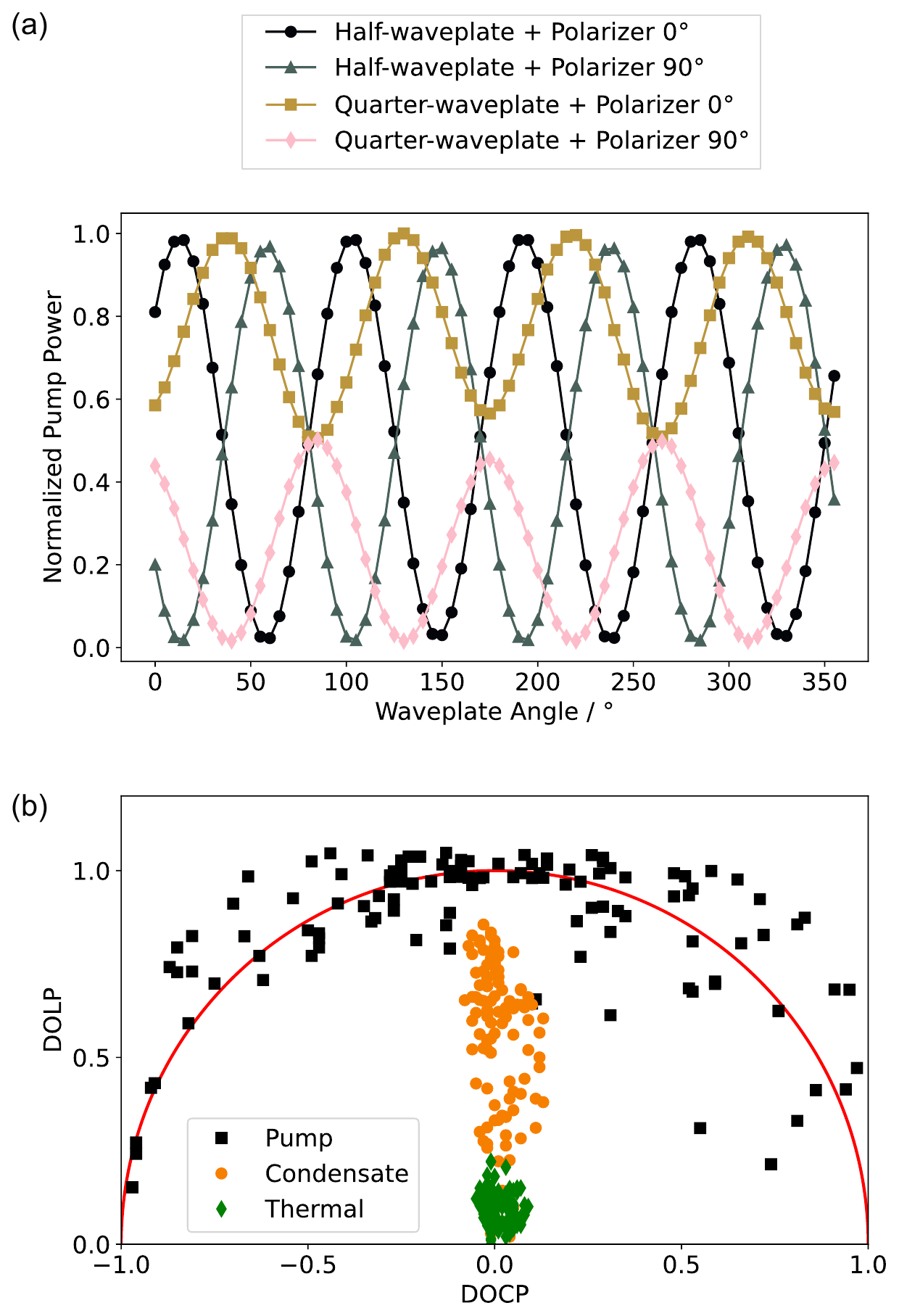}
\caption{(a) Measurement of pump laser power at different wave plate angles and two different positions of a polarizer showing the purity of the pump laser. (b) Plot of linear degree of polarization DOLP over the circular degree of polarization DOCP for pump, condensate and thermal cloud. The length of a vector from the center of origin to a certain polarization state describes its purity. The red half circle with radius $r=1$ represents fully polarized states of purity $\Pi = 1$. While the pump polarization states show strong polarization for almost all polarization states, the condensate shows strong polarization for high DOLP and lower purity for low DOLP, resulting from pumping with high DOCP. The thermal part shows only low purity.}
\label{fig:PurityRepresentation}
\end{figure}

\newpage

\end{document}